\documentclass[aps,floatfix,superscriptaddress,preprint]{revtex4}

\usepackage{amssymb,amsmath,amsfonts,latexsym,graphics,epsfig,fancybox}
\usepackage{titlesec}
\usepackage{empheq}
\usepackage{cancel}
\usepackage{bm}
\usepackage{float}
\usepackage{longtable}
\usepackage{multirow}
\usepackage{booktabs}
\usepackage{array}
\usepackage{wrapfig}
\usepackage{color}
\usepackage{slashed}
\usepackage{dsfont}
\usepackage{bbm}
\usepackage{titlesec}
\usepackage [latin1]{inputenc}

\titleformat{\section}{\large\bfseries}{\thesection}{1em}{}

\newcommand{\bea}{\begin{eqnarray}}
\newcommand{\ena}{\end{eqnarray}}
\newcommand{\be}{\begin{equation}}
\newcommand{\en}{\end{equation}}
\newcommand{\ed}{\end{document}}

\newcommand{\nn}{\nonumber\\}

\newcommand{\Tr}{\mbox{\rm{tr}}}

\begin{document}

\title{Study of the nonleptonic decay \bm{$\Xi^0_c \to \Lambda^+_c \pi^-$}
  in the covariant confined quark model} 

\author{Mikhail~A.~Ivanov}
\affiliation{Bogoliubov Laboratory of Theoretical Physics,
Joint Institute for Nuclear Research, 141980 Dubna, Russia}
\author{Valery E. Lyubovitskij}
\affiliation{Institut f\"ur Theoretische Physik, Universit\"at T\"ubingen, \\
         Kepler Center for Astro and Particle Physics, \\
         Auf der Morgenstelle 14, D-72076 T\"ubingen, Germany}
\affiliation{Departamento de F\'\i sica y Centro Cient\'\i fico
         Tecnol\'ogico de Valpara\'\i so-CCTVal, \\
         Universidad T\'ecnica Federico Santa Mar\'\i a,
             Casilla 110-V, Valpara\'\i so, Chile}
\affiliation{Millennium Institute for Subatomic Physics at
         the High-Energy Frontier (SAPHIR) of ANID, \\
         Fern\'andez Concha 700, Santiago, Chile}
\author{Zhomart Tyulemissov}
\affiliation{Bogoliubov Laboratory of Theoretical Physics,
  Joint Institute for Nuclear Research, 141980 Dubna, Russia}
\affiliation{ The Institute of Nuclear Physics, Ministry of Energy of
    the Republic of Kazakhstan, 050032 Almaty, Kazakhstan}
\affiliation{ Al-Farabi Kazakh National University, 050040 Almaty, Kazakhstan}


\begin{abstract}

The nonleptonic decay  $\Xi^0_c \to \Lambda^+_c \pi^-$  with $\Delta C=0$
is systematically studied in the framework of the covariant confined quark
model accounting for both  short and long distance effects. 
The short distance effects are induced by four topologies of external and
internal weak $W^\pm$ exchange, while long distance effects are saturated
by an inclusion of the so-called pole diagrams with an intermediate
$\frac12^+$ and $\frac12^-$ baryon resonances. The contributions from
$\frac12^+$~resonances are calculated straightforwardly by accounting
for single charmed $\Sigma^0_c$ and $\Xi^{'\,+}_c$~baryons whereas  the
contributions from $\frac12^-$~resonances are calculated by using
the well-known soft-pion theorem in the current-algebra approach.
It allows to express the parity-violating S-wave amplitude in terms
of parity-conserving matrix elements. It is found that the contribution 
of  external and internal $W$-exchange diagrams is significantly suppressed
by more than one order of magnitude in comparison with data.
The pole diagrams play the major role to get consistency with experiment.  
\end{abstract}
\maketitle


\section{Introduction}
\label{sec:intro}

The study of the heavy-flavor-conserving nonleptonic weak decays of heavy
baryons has received a lot of attention due to their observation and
measurement of branching fractions by the LHCb and Belle collaborations.
The decay $\Xi^0_c \to \Lambda^+_c + \pi^-$ was first observed at LHCb
experiment and the branching fraction was measured  to be
${\mathcal B}=(0.55\pm0.02\pm0.18)\%$~\cite{LHCb:2020gge}.
Recent experimental data obtained by the Belle collaboration  gave the value of
${\mathcal B}(\Xi^0_c \to \Lambda^+_c + \pi^-)=(0.54\pm0.05\pm0.12)\%$~\cite{Belle:2022kqi}
which is in perfect agreement with the LHCb result.

The recent theoretical review of nonleptonic two-body decays of single and
doubly charm baryons was given in Ref.~\cite{Groote:2021pxt}.
The review was aiming to shed new light on the standard current algebra
approach to such processes.

The heavy-flavor-conserving nonleptonic weak decays of heavy
baryons were studied in~\cite{Cheng:1992ff} in the formalism which incorporates
both the heavy quark symmetry and the chiral symmetry.
The branching fractions of specific nonleptonic decays such
as $\Xi_c\to\Lambda^+_c\pi$  are found to be of the order of $10^{-4}$.

The weak decays  $\Xi_b\to\Lambda_b\pi$  and $\Xi_c\to\Lambda^+_c\pi$, in which
the heavy quark is not destroyed, have been discussed in
Ref.~\cite{Voloshin:2000et}. It was shown that
these should go at the rate of order $\approx 0.01 \, \text{ps}^{-1}$.  
In the updated research~\cite{Voloshin:2019ngb} of the Voloshin's approach,
the new measurements by LHCb~\cite{LHCb:2019ldj}  of the lifetimes of
the $\Lambda_c^+$, $\Xi_c^+$ and $\Xi_c^0$ charm baryons have been used
to predict a lower bound on the rate of the decays $\Xi^0_c\to\Lambda^+_c\pi$.
It was found that ${\mathcal B}(\Xi^0_c \to \Lambda^+_c + \pi^-)>
{\mathcal B}_{\rm min}=(0.25\pm 0.15)\times 10^{-3}$.

The heavy flavor conserving decays of strange charmed baryons
proceed via two subprocesses, first, via decay $s\to u(\bar ud)$
(or equivalently, via the transition $us\to ud$), and, second,
via the transition $cs\to cd$. In Ref.~\cite{Gronau:2016xiq}
it was shown that a second term is  approximately equal
to the first term. But it was unclear whether they interfere destructively or
constructively.  For constructive interference it was found
that ${\mathcal B}(\Xi^0_c\to\Lambda^+_c+\pi^-) = (1.94 \pm 0.70) \times 10^{-3}$. 
For destructive interference, the value of  branching fraction is
expected to be less than about $10^{-4}$.

In Ref.~\cite{Faller:2015oma} the upper bound for the decay width
$\Gamma (\Xi^0_c \to \Lambda^+_c + \pi^-) < 1.7 \times 10^{-14}$ GeV was obtained
by using the  Voloshin's approach.
In work~\cite{Cheng:2015ckx} the four-quark matrix element of
heavy-flavor-conserving hadronic weak decays was evaluated in using two
different models: the MIT bag model and the diquark model. All calculations
included only S-wave amplitudes and obtained
$\mathcal{B}(\Xi^0_c \to \Lambda^+_c + \pi^-) = 1.7 \times 10^{-7}$ for MIT bag model and
$\mathcal{B}(\Xi^0_c \to \Lambda^+_c + \pi^-) = 0.87 \times 10^{-4}$ for diquark model.
In updated work~\cite{Cheng:2022kea} it was confirmed that
$\Xi_c \to \Lambda^+_c \pi$ decays are indeed dominated by the
parity-conserving transition induced from nonspectator $W$-exchange and that
they receive largest contributions from the intermediate $\Sigma^0_c$ pole
terms. Also they obtained that the asymmetry parameter $\alpha$ is positive,
of order $0.70^{+0.13}_{-0.17}$ and
$\mathcal{B}(\Xi^0_c \to \Lambda^+_c + \pi^-) = 1.76^{+0.18}_{-0.12} \times 10^{-3}$.
In \cite{Cheng:2022jbr} the wave functions from the homogeneous bag model
are adopted in order to remove the center-of-mass motion of the static bag.
The calculations have been carried out under the same framework, and it has
been shown that the matrix elements of four-quark operators are enhanced about
twice and for
$\mathcal{B}(\Xi^0_c \to \Lambda^+_c + \pi^-) = ( 7.2 \pm 0.7 ) \times 10^{-3}$. 

It was investigated pion emission and pole terms in the heavy quark conserving
weak decay of $\Xi^0_c$ in the framework of non-relativistic constituent quark
model~\cite{Niu:2021qcc}. The parity-conserving pole terms are found dominant
and the direct pion emission contributions are rather small and
$\mathcal{B}(\Xi^0_c \to \Lambda^+_c + \pi^-) = (0.58 \pm 0.21) \%$ with uncertainties
caused by the quark model parameters with 20\% errors.


This work is aiming to study the decay
$\Xi^0_c \to \Lambda^+_c + \pi^-$ in the framework of the covariant
confined quark model (CCQM) previously developed by us, see
Ref.~\cite{Branz:2009cd}. This approach found many applications,
particularly, in physics of  heavy baryons, see Refs.~\cite{Gutsche:2018msz,Ivanov:2021huf,Gutsche:2015mxa,Ivanov:2020xmw,Ivanov:2020iaq,Gutsche:2019iac,Gutsche:2019wgu,Gutsche:2018utw,Ivanov:2017axg,Gutsche:2017hux,Gutsche:2017wag,Gutsche:2015rrt,Gutsche:2015lea,Habyl:2015xka,Gutsche:2014zna,Gutsche:2013oea,Gutsche:2013pp,Gutsche:2012ze}. 
One of the important step in development of the CCQM was done in
Ref.~\cite{Gutsche:2018msz} where {\it ab initio} three-loop quark model
calculation of the $W$-exchange contribution to the nonleptonic two-body decays
of the doubly charmed baryons $\Xi_{cc}^{++}$ and $\Omega_{cc}^{+}$ have been made.
The $W$-exchange contributions appear in addition to the factorizable 
contributions and, generally,  are not suppressed. 
In \cite{Ivanov:2021huf} such an approach was extended to
study two-body nonleptonic decays of light lambda hyperon
$\Lambda \to p \pi^- (n\pi^0)$  with account for both  short and long distance
effects. It was shown that the contribution from the $W$-exchange diagrams
is sizably suppressed and basically the pole diagrams allow to describe
the experimental data for the branching fractions.

The paper is organized as follows. In Sec.~\ref{sec:single_charm}
we briefly discuss the classification and spectroscopy of singly charmed
$1/2^+$ baryons. Then we give the basic ingredients and milestones that
are needed for calculation of two-body nonleptonic decays including both
the $W$-exchange quark  and pole diagrams.  Sec.~\ref{sec:calculation} is
devoted to calculation of the matrix elements and branching fraction of the
decay $\Xi^0_c \to \Lambda^+_c + \pi^-$.
We discuss in details the classification of the diagrams appearing
in these decays and give the analytical expressions for matrix elements.
In Sec.~\ref{sec:results} we present numerical results for
the amplitudes and branching fractions. We compare them with those
available in the literature.
Finally, in Sec.~\ref{sec:summary} we make conclusions and
summarize the main results obtained in this paper.

\section{The singly charmed baryons}
\label{sec:single_charm}

The masses of singly charmed baryons have been predicted
in one gluon exchange model developed in Ref.~\cite{DeRujula:1975qlm}.
The comprehensive review on heavy baryons, their spectroscopy,
semileptonic and nonleptonic decays may be found in Ref.~\cite{Korner:1994nh}.
In Tables \ref{tab:charm_bar12}  we display the names, quark contents and
interpolating currents of the low-lying multiplets of singly charmed
baryons with spin~1/2. For singly charmed baryons the flavor decomposition
of the diquark, made of $(u,d,s)$-quarks is $3\otimes 3=\bar 3_A+ 6_S$
(A=antisymmetric, S=symmetric).  The values of masses with errors are
taken from particle data group (PDG)~\cite{ParticleDataGroup:2022pth}.
\begin{table}[H]
  \caption{Singly charmed $1/2^+$ baryon states.
    Notation $[a,b]$ and $\{a,b\}$
   for antisymmetric and symmetric flavor index combinations.}
  \centering
\vspace*{2mm}  
\def\arraystretch{1.2}
\begin{tabular}{ m{3em} m{2cm} m{2cm} m{2cm}  m{4cm}  m{3cm} }
\hline  
Title             &  Content   & $SU(3)$   & $(I,I_3)$ & Current  & Mass (MeV)
\\
\hline
$\Lambda^+_c$     &     $c[ud]$      & $\bar 3$   &   (0,0)
& $\epsilon^{abc} c^a(u^b C\gamma_5 d^c)$ & 2286.46 $\pm$ 0.14
\\
$\Xi_c^+$         &     $c[us]$      & $\bar 3$   & (1/2,1/2)
& $\epsilon^{abc} c^a(u^b C\gamma_5 s^c$) & 2467.71 $\pm$ 0.23
\\
$\Xi_c^0$         &     $c[ds]$      & $\bar 3$   & (1/2,--1/2)
& $\epsilon^{abc} c^a(d^b C\gamma_5 s^c$) & 2470.44 $\pm$ 0.28
\\
\hline
$\Sigma_c^{++}$    &     $cuu$       &  6        &   (1,1)
& $\epsilon^{abc} \gamma_\mu\gamma_5  c^a(u^b C\gamma^\mu u^c)$    & 2453.97 $\pm$ 0.14
\\
$\Sigma_c^{+}$     &    $c\{ud\}$    &  6        &   (1,0)
& $\epsilon^{abc} \gamma_\mu\gamma_5  c^a(u^b C\gamma^\mu d^c)$    & 2452.65 $\pm$ 0.22
\\
$\Sigma_c^{0}$     &    $cdd$        &  6        &  (1,--1)
& $\epsilon^{abc} \gamma_\mu\gamma_5  c^a(d^b C\gamma^\mu d^c)$   & 2453.75 $\pm$ 0.14
\\
$\Xi_c^{\prime\,+}$  & $c\{us\}$       &  6        &  (1/2,1/2)
& $\epsilon^{abc} \gamma_\mu\gamma_5  c^a(u^b C\gamma^\mu s^c)$  & 2578.2 $\pm$ 0.5
\\
$\Xi_c^{\prime\,0}$  & $c\{ds\}$       &  6        & (1/2,--1/2)
& $\epsilon^{abc} \gamma_\mu\gamma_5  c^a(d^b C\gamma^\mu s^c)$  & 2578.7 $\pm$ 0.5
\\
$\Omega_c^{0}$      & $css$          &  6        &    (0,0)
& $\epsilon^{abc} \gamma_\mu\gamma_5  c^a(s^b C\gamma^\mu s^c)$ & 2695.2 $\pm$ 1.7
\\
\hline
\end{tabular}
\label{tab:charm_bar12}
\end{table}

We are aiming to study the two-body nonleptonic decay
$\Xi^0_c \to \Lambda^+_c \pi^-$ which branching fraction was measured for
the first time by LHCb collaboration~\cite{LHCb:2020gge}.
The effective Hamiltonian relevant for this purpose is written as
\bea
\mathcal {H}^{\Delta\rm S=1}_{\rm eff} &=& \frac{G_F}{\sqrt{2}} 
\left[
\,\,  V^\ast_{us} V_{ud} \left( C^{(u)}_1(\mu_u) Q^{(u)}_1 + C^{(u)}_2(\mu_u) Q^{(u)}_2 \right)
\right.
\nn  
&&
\left.
\phantom{\frac{G_F}{\sqrt{2}}}  
+ V^\ast_{cs} V_{cd} \left( C^{(c)}_1(\mu_c) Q^{(c)}_1 + C^{(c)}_2(\mu_c) Q^{(c)}_2 \right)
+ {\rm H.c.}
\right]
\label{eq:Heff}
\ena
where $Q_1$ and $Q_2$ is the set of flavor-changing effective four-quark operators
given by
\bea
Q^{(u)}_1 &=&  (\bar s_a O_L^\mu  u_b)(\bar u_b O_{\mu L} d_a),
\qquad
Q^{(u)}_2  =   (\bar s_a O_L^\mu  u_a)(\bar u_b O_{\mu L} d_b),
\nn
Q^{(c)}_1 &=&  (\bar s_a O_L^\mu c_b)(\bar c_b O_{\mu L} d_a),
 \qquad
 Q^{(c)}_2  =   (\bar s_a O_L^\mu c_a)(\bar c_b O_{\mu L} d_b).
 \label{eq:operators}
 \ena
Here $O_L^\mu  = \gamma^\mu (1-\gamma_5 )$ is the left-handed chiral weak
matrix. One has to note that we adopt the numeration of the operators
from Ref.~\cite{Buchalla:1995vs} where the $C_2Q_2$ means the leading order
whereas  the $C_1Q_1$ is for subleading order. 
The numerical values of the Wilson coefficients $C_1$ and $C_2$  from
Ref.~\cite{Buchalla:1995vs} are being equal to
\bea
C^{(u)}_1(\mu_u) &=& -0.625, \qquad C^{(u)}_2(\mu_u) = 1.361,
\quad (\mu_u=O(1\, \text{GeV})),
\nn
C^{(c)}_1(\mu_c) &=& -0.621, \qquad C^{(c)}_2(\mu_c) = 1.336, \quad (\mu_c=O(m_c)).
\label{eq:Wilson}
 \ena
We do not include penguin operators because their Wilson coefficients
are small compare with those from current-current operators. 

In the standard model (SM) the relation  $ V_{cs}^\ast V_{cd}  = - V_{us}^\ast V_{ud} $
holds to an excellent approximation. For instance, in the Wolfenstein parametrization
of the Cabibbo-Kobayashi-Maskawa (CKM) matrix, one has
$V_{us}^\ast V_{ud}=+\lambda(1-\lambda^2) + O(\lambda^4)$
whereas  $V_{cs}^\ast V_{cd}=-\lambda(1-\lambda^2) + O(\lambda^4).$
The global fit in the SM for the Wolfenstein parameter
gives $\lambda= 0.22500\pm 0.00067$.   
In what follows, we introduce the short notations
\be
V_{\rm CKM}^{(u)} =   |V_{us}^\ast V_{ud}|, \quad\text{and}\quad
V_{\rm CKM}^{(c)} =  -|V_{cs}^\ast V_{cd}|.
\en
The numerical values of the CKM matrix elements needed in our calculations are
taken from PDG~\cite{ParticleDataGroup:2022pth}:
\begin{align}
|V_{ud}| &= 0.97373\pm0.00031, & |V_{us}| &= 0.2243\pm0.0008,
\nn
|V_{cd}| &= 0.221\pm0.004,     & |V_{cs}| &= 0.975\pm0.006,
\end{align}
that approximately give $V_{\rm CKM}^{(u)} \approx  0.218$
and  $V_{\rm CKM}^{(c)} \approx  -0.215$.

\begin{figure}[H]
 \centering
\includegraphics[width=0.7\textwidth]{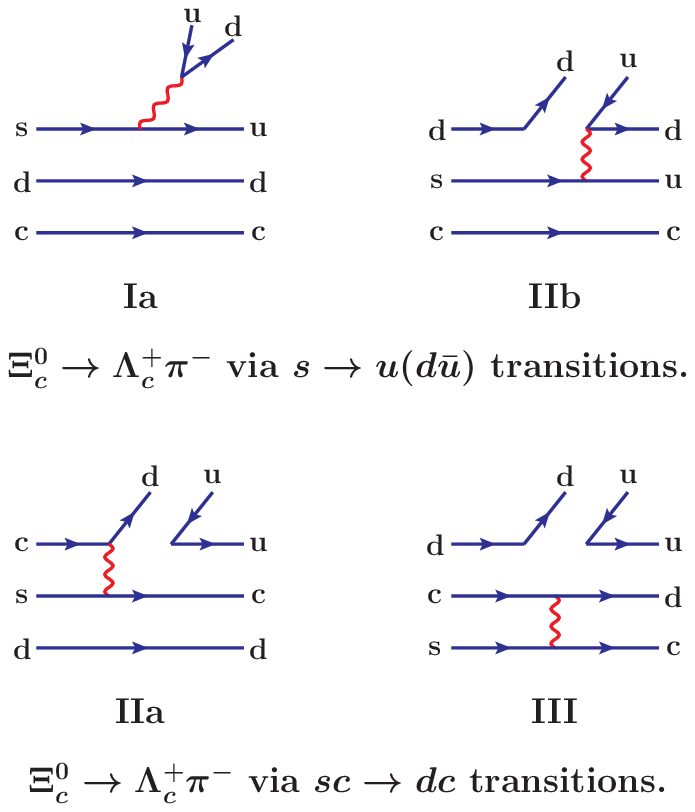} 
\caption{ Flavor-color topologies for $\Xi^0_c\to\Lambda_c^+\pi^-$ decay:
  Ia is the tree level diagram, IIa, IIb and III are
  the $W$-exchange diagrams. }
\label{fig:tpg}
 \end{figure}

\begin{figure}[H]
\centering
\includegraphics[width=0.9\textwidth]{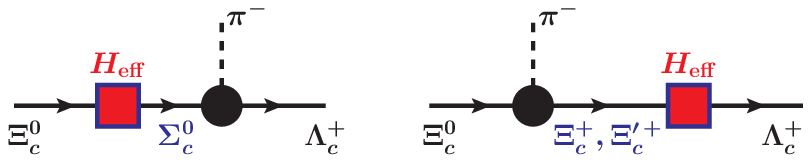}
\caption{The pole diagrams which effectively account for
the long-distance contributions.}
\label{fig:pole}
\end{figure}

The quark diagrams that contribute to the Cabibbo-favored decay
are shown in Fig.~\ref{fig:tpg}. After hadronizarion,
the diagram Ia factorizes out into two parts: the weak transition
$\Xi_c^0\to \Lambda_c^+$ via the $W$-emission and the matrix element
describing the pion leptonic decay. The $W$-exchange diagrams IIa, IIb and
III contribute into both the pure quark diagrams called the short
distance (SD) contributions and effectively into the pole diagrams
shown in Fig.~\ref{fig:pole}. They describe the so-called long distance
(LD)  contributions. For instance, the diagrams IIa and III effectively
generate the $\Sigma^0_c$-resonance diagram, whereas the diagram IIb
effectively generates the $\Xi_c^+$ and $\Xi_c^{\prime\,+}$-resonance diagrams. 


\section{Matrix elements and decay widths}
\label{sec:calculation}

We are going to calculate the matrix elements of nonleptonic decays of
$\Xi^0_c$-baryon in the framework of the  CCQM developed in our previous papers.
The starting point is the Lagrangian describing couplings of the baryon field
with its interpolating quark current.
 \be
    {\cal L}(x) = g_B \bar B(x) \, J(x) + {\rm H.c.}
     \label{eq:Lag} \,,
 \en   
 where the coupling constant $g_B$ is determined from the so-called 
 {\it compositeness condition}, which was proposed by Salam  and
 Weinberg~\cite{Salam:1962ap,Weinberg:1962hj}
 and extensively used in the literature (see, e.g.,
 Refs.~\cite{Hayashi:1967hk,Efimov:1993ei}). 
 
 The nonlocal extension of the interpolating currents shown in
 Table~\ref{tab:charm_bar12}  reads
 \bea
 J_B(x)  &=& \int\!\! dx_1 \!\! \int\!\! dx_2 \!\! \int\!\! dx_3 \,
 F_B(x;x_1,x_2,x_3) \,
 \varepsilon_{abc}\,\Gamma_1 q_1^a(x_1)\,
 \left(q_2^b(x_2) \,C\Gamma_2\, q_3^c(x_3)\right)\,,
 \nn
 F_B(x;x_1,x_2,x_3) &=& \delta^{(4)}\Big(x-\sum\limits_{i=1}^3 w_i x_i\Big)
 \Phi_B\Big(\sum\limits_{i<j}(x_i-x_j)^2\Big) \,,
 \label{eq:nonlocal-cur}
 \ena
 where $w_i=m_i/(\sum_{j=1}^3 m_j)$ and $m_i$ is the mass of the quark
 at the space-time point $x_i$. The matrices $\Gamma_1,\Gamma_2$ are the Dirac
 strings of the initial and final baryon states
 as specified  in Table~\ref{tab:charm_bar12}.
 The vertex function $\Phi_B$ is written as
 \bea
 &&
 \Phi_B\Big(\sum\limits_{i<j}(x_i-x_j)^2\Big) =
 \int\!\frac{dq_1}{(2\pi)^4}\int\!\frac{dq_2}{(2\pi)^4}
 e^{-iq_1(x_1-x_3)-iq_2(x_2-x_3)}
 \widetilde\Phi_B\Big(-\vec\Omega_q^2\Big)\,,
 \nn
 &&
 \widetilde\Phi_B\Big(-\vec\Omega_q^2\Big) =
 \exp\left(\vec\Omega_q^2/\Lambda_B^2\right), \qquad
 \vec\Omega_q^2=\tfrac12 (q_1+q_2)^2 + \tfrac16 (q_1-q_2)^2 =
 \frac23\sum\limits_{i\le j} q_iq_j\,.
 \label{eq:vertex}
 \ena
 For simplicity and calculational advantages we mostly
 adopted a Gaussian form for the  functions $\widetilde\Phi_B$.
 Here $\Lambda_B$ is the size parameter for a given
 baryon. The size parameter phenomenologically describes 
 the distribution of the constituent quarks
 in the given baryon. 

\begin{figure}[H]
\centering
\includegraphics[width=0.7\textwidth]{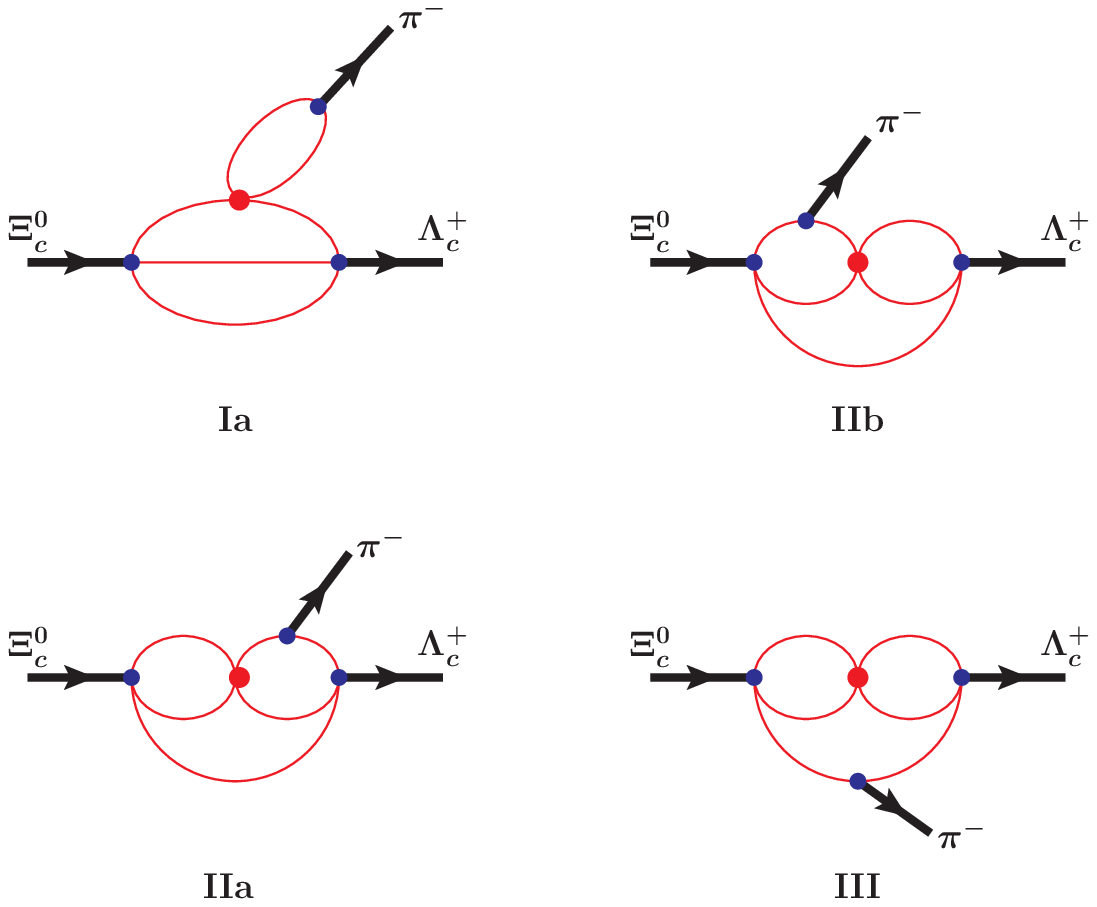}
\caption{Quark diagrams describing the SD-contributions}
\label{fig:SD}
\end{figure}

\begin{figure}[H]
\centering
\includegraphics[width=0.8\textwidth]{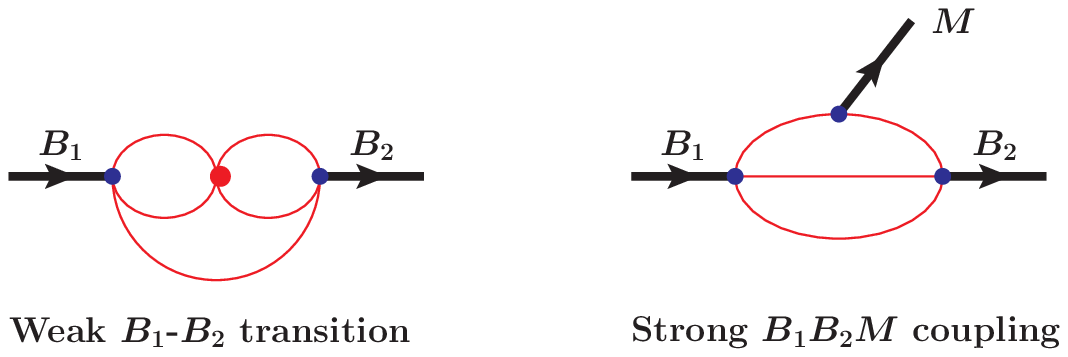}
\caption{Feynman diagrams describing the building blocks
  of the LD-contributions}
\label{fig:LD-block}
\end{figure}

 In our approach the matrix elements contributing to the baryon
 transitions $\Xi^0_c\to\Lambda_c^+\pi^-$ are represented by
 a set of the quark diagrams shown in Fig.~\ref{fig:SD}.
 They describe the so-called short distance contributions.
The diagrams describing the building blocks of the LD-contributions
are shown in Fig.~\ref{fig:LD-block}.
First, we discuss the matrix elements corresponding to the SD-contributions.
One has
\bea  
M_{\rm SD} (\Xi^0_c \to\Lambda_c^+\pi^-) &=&
\frac{G_F}{\sqrt{2}}\Big\{ V_{\rm CKM}^{(u)}\,
\bar u(p_2)\Big[ (C_2^{(u)} +\xi C_1^{(u)})\,D_{\rm Ia}
  + (C_2^{(u)}-C_1^{(u)})\,D_{\rm IIb}\Big] u(p_1)
\nn
&&
\phantom{\frac{G_F}{\sqrt{2}}}
+ V_{\rm CKM}^{(c)} (C_2^{(c)}-C_1^{(c)})\,
\bar u(p_2)\Big[ D_{\rm IIa} + D_{\rm III}\Big] u(p_1)
\Big\}\,.
\label{eq-SD:matrix_element}
\ena
Here, the factor $\xi=1/N_c$ where $N_c$ is the number of colors.
This factor is set to zero in the numerical calculations
according to the widely accepted phenomenology of the nonleptonic decays.

The contribution from the tree diagram factorizes into two pieces
\bea
D_{\rm Ia}  &=&
N_c\,g_{M}\int\!\!\frac{d^4k}{(2\pi)^4i}\widetilde\Phi_{M}(-k^2)\,
\Tr\left[O^\mu_L S_u(k-w_u q) \gamma_5 S_d(k+w_d q)\right]
\nn
&\times&
6\,g_{B_1}g_{B_2}\int\!\!\frac{d^4k_1}{(2\pi)^4i}\int\!\!\frac{d^4k_2}{(2\pi)^4i}
\widetilde\Phi_{B_1}\Big(-\vec\Omega_q^{\,2}\Big)
\widetilde\Phi_{B_2}\Big(-\vec\Omega_r^{\,2}\Big)
\nn
&&
\times S_c(k_2)\Tr\left[ S_u(k_1+p_2) O_{\mu\, L} S_s(k_1+p_1)\gamma_5
  S_d(k_1+k_2)\gamma_5\right]
\nn[1.2ex] 
&=&
- 6 f_M q^\mu g_{B_1}g_{B_2}
\int\!\!\frac{d^4k_1}{(2\pi)^4i}\int\!\!\frac{d^4k_2}{(2\pi)^4i}
\widetilde\Phi_{B_1}\Big(-\vec\Omega_q^{\,2}\Big)
\widetilde\Phi_{B_2}\Big(-\vec\Omega_r^{\,2}\Big)
\nn
&&
\times S_c(k_2)\Tr\left[ S_u(k_1+p_2) O_{\mu\, L} S_s(k_1+p_1)\gamma_5
  S_d(k_1+k_2)\gamma_5\right]
\label{eq:tree}
\ena
where $q_1=k_2-w_1^{\rm in} p_1$, $q_2=-k_1-k_2-w_2^{\rm in} p_1$ and
$r_1=-k_2+w_1^{\rm out} p_2$, $r_2=-k_1-p_2+w_2^{\rm out} p_2$. The expression
for $\vec\Omega^{\,2}$ is given by Eq.~(\ref{eq:vertex}).
Hereafter we adopt the brief notations $B_1$ for the ingoing baryon with
the momentum $p_1$, $B_2$ for the outgoing baryon with
the momentum $p_2$ and $M$ for   the outgoing meson with the momentum $q$.
The minus sign  in front of $f_M$ appears because the momentum $q$ flows
in the opposite direction from the decay of $M$-meson.

The calculation of the three-loop $W$--exchange diagrams
is much more involved because the matrix element does not factorize.
One has
\bea
D_{\rm IIb} &=&
12 g_{B_1}g_{B_2}g_{M}
\Big[ \prod\limits_{i=1}^3\int\!\!\frac{d^4k_i}{(2\pi)^4i} \Big]
\widetilde\Phi_{B_1}\Big(-\vec\Omega_q^{\,2}\Big)
\widetilde\Phi_{B_2}\Big(-\vec\Omega_r^{\,2}\Big)
\widetilde\Phi_{M}(-P^2)\,
\nn
&\times&
S_c(k_3)\Tr\left[\gamma_5 S_d(k_2+p_2)(1+\gamma_5)S_u(k_2+k_3)\right]
\nn
&\times&
\Tr\left[S_u(k_1+p_2)\gamma_5S_d(k_1+p_1)\gamma_5 S_s(k_1+k_3)(1-\gamma_5)\right],
\nn[1.2ex]
&& q_1 = k_3-w_1^{\rm in} p_1, \quad  q_2 = k_1+p_1-w_2^{\rm in} p_1,
\nn
&& r_1 = -k_3+w_1^{\rm out} p_2, \quad  r_2 = k_2+k_3+w_2^{\rm out} p_2,\qquad
P=k_1+w_u p_1 + w_d p_2.
\label{eq:IIb}
\ena
\bea
D_{\rm IIa} &=&
6 g_{B_1}g_{B_2}g_{M}
\Big[ \prod\limits_{i=1}^3\int\!\!\frac{d^4k_i}{(2\pi)^4i} \Big]
\widetilde\Phi_{B_1}\Big(-\vec\Omega_q^{\,2}\Big)
\widetilde\Phi_{B_2}\Big(-\vec\Omega_r^{\,2}\Big)
\widetilde\Phi_{M}(-P^2)\,
\nn
&\times&
S_c(k_3-k_2)O^\mu_LS_c(k_1+p_1)
\Tr\left[S_d(k_3)\gamma_5 S_u(k_2+p_2)\gamma_5 S_d(k_2+p_1)O_{\mu\, L}
  S_s(k_3-k_1)\gamma_5\right]
\nn[1.2ex]
&& q_1 = k_1+p_1-w_1^{\rm in} p_1, \quad  q_2 = -k_3-w_2^{\rm in} p_1,
\nn
&& r_1 = k_2-k_3+w_1^{\rm out} p_2, \quad  r_2 = -k_2-p_2+w_2^{\rm out} p_2,\qquad
P=k_2+w_u p_1 + w_d p_2.
\label{eq:IIa}
\ena
\bea
D_{\rm III} &=&
6 g_{B_1}g_{B_2}g_{M}
\Big[\prod\limits_{i=1}^3\int\!\!\frac{d^4k_i}{(2\pi)^4i} \Big]
\widetilde\Phi_{B_1}\Big(-\vec\Omega_q^{\,2}\Big)
\widetilde\Phi_{B_2}\Big(-\vec\Omega_r^{\,2}\Big)
\widetilde\Phi_{M}(-P^2)\,
\nn
&\times&
S_c(k_3)O^\mu_LS_c(k_2)
\Tr\left[S_d(k_1+k_3)\gamma_5 S_u(k_1+p_2)\gamma_5 S_d(k_1+p_1)\gamma_5
  S_s(k_1+k_2)O_{\mu\, R}\right]
\nn[1.2ex]
&& q_1 = k_2-w_1^{\rm in} p_1, \quad  q_2 = k_1+p_1-w_2^{\rm in} p_1,
\nn
&& r_1 = -k_3+w_1^{\rm out} p_2, \quad  r_2 = -k_1-p_2+w_2^{\rm out} p_2,\qquad
P=k_1+w_u p_1 + w_d p_2.
\label{eq:III}
\ena
The calculation of the three-loop integrals proceeds in two steps, first,
one has to perform the loop integration by using  Fock-Schwinger representation
for the quark propagators and Gaussian form for the vertex functions.
This allows  one to do tensor loop integrals in a very efficient way since
one can convert loop momenta into derivatives of the exponent function. 
The calculations are done by using a FORM code which
works for any numbers of loops and propagators.
Second, one has to calculate the obtained integrals numerically
over Fock-Schwinger variables by adopting the quark confinement anzatz.
The numerical calculations are done by using the FORTRAN codes
which include the output from the FORM code written in the format
of  double precision accuracy. 
Since the files with  the output from FORM contain several thousand lines
we are unable to show them in the paper.
The details of such calculations may be found in our recent
papers~\cite{Gutsche:2017hux,Gutsche:2018utw}. 
The calculation is quite time consuming both analytically and numerically.

Finally, the matrix element describing the SD-contributions are written
as
\be  
M_{\rm SD} (\Xi^0_c \to\Lambda_c^+\pi^-) =
\frac{G_F}{\sqrt{2}}
\bar u(p_2)\Big( A_{\rm\, SD} + \gamma_5 B_{\rm\, SD}\Big)u(p_1),
\label{eq-SD:full-matrix_element}
\en
where
\bea
A_{\rm\, SD} &=&  V_{\rm CKM}^{(u)}
\Big[ (C_2^{(u)} +\xi C_1^{(u)}) a_{\rm Ia} + (C_2^{(u)}-C_1^{(u)}) a_{\rm IIb}\Big]
 + V_{\rm CKM}^{(c)} (C_2^{(c)}-C_1^{(c)}) \Big( a_{\rm IIa} + a_{\rm III}\Big),
 \nn
B_{\rm\, SD} &=&  V_{\rm CKM}^{(u)}
\Big[ (C_2^{(u)} +\xi C_1^{(u)})\,b_{\rm Ia} + (C_2^{(u)}-C_1^{(u)})\,b_{\rm IIb}\Big]
 + V_{\rm CKM}^{(c)} (C_2^{(c)}-C_1^{(c)})\Big( b_{\rm IIa} + b_{\rm III}\Big) .
\nonumber
 \ena



Now, we discuss the matrix elements corresponding to the LD contributions.
The contribution coming from the pole diagram in Fig.~\ref{fig:pole}
with the $\Sigma^0_c$-resonance is written as 
\be 
M_{\,\Sigma^0_c} =
\frac{G_F}{\sqrt{2}} V_{\rm CKM}^{(c)} \left(C_1^{(c)}-C_2^{(c)}\right)
\bar u(p_2) D_{\,\Sigma^0_c\,\Lambda_c^+\,\pi^-}(p_1,p_2) S_{\,\Sigma^0_c}(p_1)
D_{\,\Xi^0_c\,\Sigma^0_c}(p_1) u(p_1)
\label{eq-LD-I:matrix_element}
\en
where $ S_{\,\Sigma^0_c}(p_1) = 1/(m_{\Sigma^0_c}-\not\!{p}_1) $.
The explicit form of $D$-functions are written down as
\bea
D_{\,\Sigma^0_c\,\Lambda_c^+\,\pi^-} &=&
12\, g_{\,\Sigma^0_c}\, g_{\,\Lambda_c^+}\, g_{\,\pi^-}
\Big[ \prod\limits_{i=1}^2\int\!\!\frac{d^4k_i}{(2\pi)^4i} \Big]
\widetilde\Phi_{\,\Sigma^0_c}\Big(-\vec\Omega_q^{\,2}\Big)
\widetilde\Phi_{\,\Lambda_c^+}\Big(-\vec\Omega_r^{\,2}\Big)
\widetilde\Phi_{\,\pi^-}(-P^2)\,
\nn
&\times&
S_c(k_2)\gamma_\alpha\gamma_5\, \Tr\left[
  \gamma_5 S_u(k_1+p_2)\gamma_5 S_d(k_1+p_1)\gamma^\alpha S_d(k_1+k_2) \right]
\ena
where $q_1  = k_2-w_1^{\rm res} p_1, q_2 = -k_1-k_2-w_2^{\rm res} p_1$,
$ r_1 =-k_2+w_1^{\rm out} p_2, r_2 = -k_1-(1-w_2^{\rm out}) p_2,$
and $P=k_1+w_1^{M} p_1 + w_2^{M} p_2$. Here the notations are
${\rm ``res''}=\Sigma^0_c$, $ {\rm ``out''} = \Lambda_c^+$ and $M=\pi^-$.

\bea
D_{\,\Xi^0_c\,\Sigma^0_c} &=&
12\, g_{\,\Xi^0_c}\,g_{\,\Sigma^0_c}
\Big[ \prod\limits_{i=1}^3\int\!\!\frac{d^4k_i}{(2\pi)^4i} \Big]
\widetilde\Phi_{\,\Xi^0_c}\Big(-\vec\Omega_q^{\,2}\Big)
\widetilde\Phi_{\,\Sigma^0_c}\Big(-\vec\Omega_r^{\,2}\Big)
\nn
&\times&
\gamma_\alpha\gamma_5 S_c(k_2+p_1)O_{\mu\,L} S_c(k_1+p_1)\,\Tr\left[
 S_d(k_3)\gamma^\alpha S_d(k_3-k_2) O^\mu_L S_s(k_3-k_1)\gamma_5 \right]
\ena
where $q_1 = k_1+(1-w_1^{\rm in}) p_1,\, q_2 = -k_3-w_2^{\rm in} p_1$,
$r_1  =-k_2-(1-w_1^{\rm res}) p_1, \,  r_2 = k_3+w_2^{\rm res} p_1$ and
 $ {\rm ``in''} = \Xi^0_c$.

By using the mass-shell conditions, one obtains
\bea
\bar u(p_2)D_{\,\Sigma^0_c\,\Lambda_c^+\,\pi^-}(p_1,p_2) &=&
\bar u(p_2)\gamma_5(g^{(0)}_{\,\Sigma^0_c\,\Lambda_c^+\,\pi^-}
+\not\!{p}_1 \, g^{(1)}_{\,\Sigma^0_c\,\Lambda_c^+\,\pi^-}),
\nn
D_{\,\Xi^0_c\,\Sigma^0_c}(p_1) u(p_1) &=&
(a_{\,\Xi^0_c\,\Sigma^0_c}+\gamma_5\,b_{\,\Xi^0_c\,\Sigma^0_c})u(p_1).
\label{eq:B1R1}
\ena
The final expression for the $\Sigma^0_c$-resonance  diagram is written as
\be
M_{\Sigma^0_c} =  \frac{G_F}{\sqrt{2}}\,
\bar u(p_2) \Big(A_{\Sigma^0_c} + \gamma_5\, B_{\Sigma^0_c} \Big) u(p_1),
\label{eq:LD-I}
\en
where
\bea
A_{\Sigma^0_c} &=&  V_{\rm CKM}^{(c)}\,\left(C_1^{(c)}-C_2^{(c)}\right)\,
\frac{ g^{(-)}_{\,\Sigma^0_c\,\Lambda_c^+\,\pi^-} b_{\,\Xi^0_c\,\Sigma^0_c}}
     {m_{\Sigma^0_c} + m_{\Xi^0_c}}, \,\,\,\text{where}\,\,\,
g^{(-)}_{\,\Sigma^0_c\,\Lambda_c^+\,\pi^-} = g^{(0)}_{\,\Sigma^0_c\,\Lambda_c^+\,\pi^-}
- m_{\Xi^0_c} \, g^{(1)}_{\,\Sigma^0_c\,\Lambda_c^+\,\pi^-} ,
\nn
B_{\Sigma^0_c} &=&  V_{\rm CKM}^{(c)}\,\left(C_1^{(c)}-C_2^{(c)}\right)\,
\frac{ g^{(+)}_{\,\Sigma^0_c\,\Lambda_c^+\,\pi^-} a_{\,\Xi^0_c\,\Sigma^0_c}}
     {m_{\Sigma^0_c} - m_{\Xi^0_c}}, \,\,\,\text{where}\,\,\,
g^{(+)}_{\,\Sigma^0_c\,\Lambda_c^+\,\pi^-} = g^{(0)}_{\,\Sigma^0_c\,\Lambda_c^+\,\pi^-}
+ m_{\Xi^0_c} \, g^{(1)}_{\,\Sigma^0_c\,\Lambda_c^+\,\pi^-}.
\nonumber
\ena


The matrix elements corresponding to the LD-contributions
coming from the second diagram in Fig.~\ref{fig:pole} with
$B_{\rm res}=\Xi^+_c,\Xi^{'\,+}_c$ are calculated in a similar way.
We perform the necessary steps below.
\bea  
M_{\Xi^{'\,+}_c}  &=&
\frac{G_F}{\sqrt{2}}\,
\bar u(p_2)\Big\{
\left[
  V_{\rm CKM}^{(u)} \left(C_1^{(u)}-C_2^{(u)}\right) D^{(u)}_{\,\Xi^{'\,+}_c\,\Lambda^+_c}(p_2)
  + V_{\rm CKM}^{(c)} \left(C_1^{(c)}-C_2^{(c)}\right)
  D^{(c)}_{\,\Xi^{'\,+}_c\,\Lambda^+_c}(p_2)
\right]
\nn
&&\phantom{\frac{G_F}{\sqrt{2}}\,\bar u(p_2)}
\times S_{\,\Xi^{'\,+}_c}(p_2)D_{\,\Xi^0_c\,\Xi^{'\,+}_c\,\pi^-}(p_1,p_2) 
\Big\}\,u(p_1)
\label{eq-LD-II:matrix_element}
\ena
where
\bea
D^{(u)}_{\,\Xi^{'\,+}_c\,\Lambda^+_c}(p_2) &=&
-12\,g_{\,\Xi^{'\,+}_c}\, g_{\,\Lambda^+_c}
\Big[ \prod\limits_{i=1}^3\int\!\!\frac{d^4k_i}{(2\pi)^4i} \Big]
\widetilde\Phi_{\,\Xi^{'\,+}_c}\Big(-\vec\Omega_q^{\,2}\Big)
\widetilde\Phi_{\,\Lambda^+_c}\Big(-\vec\Omega_r^{\,2}\Big)
\nn
&\times&
S_c(k_3)\gamma_\alpha\gamma_5
\Tr\left[ S_u(k_2+p_2)(1+\gamma_5)S_d(k_2+k_3) \gamma_5\right]
\nn
&\times&
\Tr\left[ S_u(k_1+p_2)\gamma^\alpha S_s(k_1+k_3)(1-\gamma_5) \right]
\ena
where $q_1  = k_3-w_1^{\rm res} p_2, \,  q_2 = k_1+(1-w_2^{\rm res}) p_2$,
$r_1  =-k_3+w_1^{\rm out} p_2, \,  r_2 = -k_2-(1-w_2^{\rm out}) p_2$ and
${\rm ``res''} = \Xi^{'\,+}_c$, ${\rm ``out''} =\Lambda^+_c $. 
\bea
D^{(c)}_{\,\Xi^{'\,+}_c\,\Lambda^+_c}(p_2) &=&
-6\,g_{\,\Xi^{'\,+}_c}\, g_{\,\Lambda^+_c}
\Big[ \prod\limits_{i=1}^3\int\!\!\frac{d^4k_i}{(2\pi)^4i} \Big]
\widetilde\Phi_{\,\Xi^{'\,+}_c}\Big(-\vec\Omega_q^{\,2}\Big)
\widetilde\Phi_{\,\Lambda^+_c}\Big(-\vec\Omega_r^{\,2}\Big)
\nn
&\times&
S_c(k_2+p_2)O_{\mu\,L} S_c(k_1+p_2)\,\gamma_\alpha\gamma_5
\nn
&\times&
\Tr\left[S_u(k_3)\gamma_5 S_d(k_3-k_2) O_L^\mu S_s(k_3-k_1)\gamma^\alpha \right]
\ena
where $q_1  = k_1+(1-w_1^{\rm res}) p_2, \, q_2 = -k_3-w_2^{\rm res} p_2$,
$ r_1 = -k_2-(1-w_1^{\rm out}) p_2, \,  r_2 = k_3+w_2^{\rm out} p_2$.
\bea
D_{\,\Xi_c^0\,\Xi^{'\,+}_c\,\pi^-} &=&
-6\,g_{\,\Xi_c^0}\,g_{\,\Xi^{'\,+}_c}\,g_{\,\pi^-}
\Big[ \prod\limits_{i=1}^2\int\!\!\frac{d^4k_i}{(2\pi)^4i} \Big]
\widetilde\Phi_{\,\Xi_c^0} \Big(-\vec\Omega_q^{\,2}\Big)
\widetilde\Phi_{\,\Xi^{'\,+}_c}\Big(-\vec\Omega_r^{\,2}\Big)
\widetilde\Phi_{\,\pi^-}(-P^2)\,
\nn
&\times&
\gamma_\alpha\gamma_5\,S_c(k_2) \Tr\left[
  S_u(k_1+p_2)\gamma_5 S_d(k_1+p_1)\gamma_5 S_s(k_1+k_2)\gamma^\alpha
  \right]
\ena
where $q_1 = k_2-w_1^{\rm in} p_1, \, q_2 = k_1+(1-w_2^{\rm in}) p_1$,
$ r_1  =-k_2+w_1^{\rm res} p_2, \quad  r_2 = -k_1-(1-w_2^{\rm res}) p_2$,
$P=k_1+w_1^M p_1 + w_2^M p_2$ and 
${\rm ``in''} =\Xi^0_c $.
By using the mass-shell conditions, one obtains
\bea
\bar u(p_2) D^{(u,c)}_{\,\Xi^{'\,+}_c\,\Lambda^+_c}(p_2) &=&
\bar u(p_2) \left( a^{(u,c)}_{\,\Xi^{'\,+}_c\,\Lambda^+_c}+\gamma_5\,
             b^{(u,c)}_{\,\Xi^{'\,+}_c\,\Lambda^+_c} \right)
\nn
D_{\,\Xi^0_c\,\Xi^{'\,+}_c\,\pi^-}(p_1,p_2)u(p_1) &=&
\gamma_5 \left( g^{(0)}_{\,\Xi^0_c\,\Xi^{'\,+}_c\,\pi^-}
               + \not\!{p}_2 g^{(1)}_{\,\Xi^0_c\,\Xi^{'\,+}_c\,\pi^-} \right) u(p_1).
\nonumber
\ena
The final expression for the second pole diagram is written as
\be
M_{\Xi_c^{'\,+}} =  \frac{G_F}{\sqrt{2}}\,
\bar u(p_2) \Big(A_{\Xi_c^{'\,+}} + \gamma_5\, B_{\Xi_c^{'\,+}} \Big) u(p_1),
\label{eq:LD-II}
\en
where
\bea
A_{\Xi_c^{'\,+}} &=&
\Big\{
V_{\rm CKM}^{(u)}\,\left(C_1^{(u)}-C_2^{(u)}\right)\,b^{(u)}_{\,\Xi_c^{'\,+}\,\Lambda^+_c}
+
V_{\rm CKM}^{(c)}\,\left(C_1^{(c)}-C_2^{(c)}\right)\,b^{(c)}_{\,\Xi_c^{'\,+}\,\Lambda^+_c}
\Big\}
\frac{g^{(+)}_{\,\Xi^0_c\,\Xi^{'\,+}_c\,\pi^-} }{m_{\,\Xi_c^{'\,+}} + m_{\,\Lambda^+_c}},
\nn
B_{\Xi_c^{'\,+}} &=&
\Big\{
V_{\rm CKM}^{(u)}\,\left(C_1^{(u)}-C_2^{(u)}\right)\,a^{(u)}_{\,\Xi_c^{'\,+}\,\Lambda^+_c}
+
V_{\rm CKM}^{(c)}\,\left(C_1^{(c)}-C_2^{(c)}\right)\,a^{(c)}_{\,\Xi_c^{'\,+}\,\Lambda^+_c}
\Big\}
\frac{ g^{(-)}_{\,\Xi^0_c\,\Xi^{'\,+}_c\,\pi^-}}{m_{\,\Xi_c^{'\,+}} - m_{\,\Lambda^+_c}},
\nonumber
\ena
where $g^{(\pm)}_{\,\Xi^0_c\,\Xi^{'\,+}_c\,\pi^-}=
       g^{(0)}_{\,\Xi^0_c\,\Xi^{'\,+}_c\,\pi^-} \pm m_{\Lambda_c^+} g^{(1)}_{\,\Xi^0_c\,\Xi^{'\,+}_c\,\pi^-} $. 

\bea  
M_{\Xi^+_c}  &=&
\frac{G_F}{\sqrt{2}}\,
\bar u(p_2)\Big\{
\left[
  V_{\rm CKM}^{(u)} \left(C_2^{(u)}-C_1^{(u)}\right) D^{(u)}_{\,\Xi^{+}_c\,\Lambda^+_c}(p_2)
+ V_{\rm CKM}^{(c)} \left(C_2^{(c)}-C_1^{(c)}\right) D^{(c)}_{\,\Xi^{+}_c\,\Lambda^+_c}(p_2)
\right]
\nn
&&\phantom{\frac{G_F}{\sqrt{2}}\,\bar u(p_2)}
\times S_{\,\Xi^{+}_c}(p_2)D_{\,\Xi^0_c\,\Xi^{+}_c\,\pi^-}(p_1,p_2) 
\Big\}\,u(p_1)
\label{eq-LD-III:matrix_element}
\ena
where
\bea
D^{(u)}_{\,\Xi^{+}_c\,\Lambda^+_c}(p_2) &=&
12\,g_{\,\Xi^{+}_c}\, g_{\,\Lambda^+_c}
\Big[ \prod\limits_{i=1}^3\int\!\!\frac{d^4k_i}{(2\pi)^4i} \Big]
\widetilde\Phi_{\,\Xi^{+}_c}\Big(-\vec\Omega_q^{\,2}\Big)
\widetilde\Phi_{\,\Lambda^+_c}\Big(-\vec\Omega_r^{\,2}\Big)
\nn
&\times&
S_c(k_3)
\Tr\left[ S_u(k_2+p_2)(1+\gamma_5)S_d(k_2+k_3) \gamma_5\right]
\nn
&\times&
\Tr\left[ S_u(k_1+p_2)\gamma_5 S_s(k_1+k_3)(1-\gamma_5) \right]
\label{eq:Xicp-weak-u}
\ena
where $q_1  = k_3-w_1^{\rm res} p_2, \,  q_2 = k_1+(1-w_2^{\rm res}) p_2$,
$r_1  =-k_3+w_1^{\rm out} p_2, \,  r_2 = -k_2-(1-w_2^{\rm out}) p_2$ and
${\rm ``res''} = \Xi^{+}_c$, ${\rm ``out''} =\Lambda^+_c $. 
\bea
D^{(c)}_{\,\Xi^{+}_c\,\Lambda^+_c}(p_2) &=&
-6\,g_{\,\Xi^{+}_c}\, g_{\,\Lambda^+_c}
\Big[ \prod\limits_{i=1}^3\int\!\!\frac{d^4k_i}{(2\pi)^4i} \Big]
\widetilde\Phi_{\,\Xi^{+}_c}\Big(-\vec\Omega_q^{\,2}\Big)
\widetilde\Phi_{\,\Lambda^+_c}\Big(-\vec\Omega_r^{\,2}\Big)
\nn
&\times&
S_c(k_2+p_2)O_{\mu\,L} S_c(k_1+p_2)
\nn
&\times&
\Tr\left[S_u(k_3)\gamma_5 S_d(k_3-k_2) O_L^\mu S_s(k_3-k_1)\gamma_5\right]
\label{eq:Xicp-weak-c}
\ena
where $q_1  = k_1+(1-w_1^{\rm res}) p_2, \, q_2 = -k_3-w_2^{\rm res} p_2$,
$ r_1 = -k_2-(1-w_1^{\rm out}) p_2, \,  r_2 = k_3+w_2^{\rm out} p_2$.
By using the mass-shell conditions, one obtains
\bea
\bar u(p_2) D^{(u,c)}_{\,\Xi^{+}_c\,\Lambda^+_c}(p_2) &=&
\bar u(p_2)\left(
a^{(u,c)}_{\,\Xi^{+}_c\,\Lambda^+_c}+\gamma_5\,b^{(u,c)}_{\,\Xi^{+}_c\,\Lambda^+_c}\right)
\label{eq:Xicp-weak}
\ena             

It appears that the strong transition $\Xi_c^0\to\Xi^{+}_c+\pi^-$ is
identically equal to zero due to the chosen form of the interpolating quark
current as shown in Table~\ref{tab:charm_bar12}:
$\epsilon^{abc} c^a(u^b C\gamma_5 s^c)$. As a result, this transition is
described by the diagram which contains the trace
of a string with three quark propagators and three $\gamma_5$ matrices
that gives zero contribution.
Explicitly we have 
 \bea
D_{\,\Xi_c^0\,\Xi^{+}_c\,\pi^-} &=&
6\,g_{\,\Xi_c^0}\,g_{\,\Xi^{+}_c}\,g_{\,\pi^-}
\Big[ \prod\limits_{i=1}^2\int\!\!\frac{d^4k_i}{(2\pi)^4i} \Big]
\widetilde\Phi_{\,\Xi_c^0} \Big(-\vec\Omega_q^{\,2}\Big)
\widetilde\Phi_{\,\Xi^{+}_c}\Big(-\vec\Omega_r^{\,2}\Big)
\widetilde\Phi_{\,\pi^-}(-P^2)\,
\nn
&\times&
S_c(k_2) \Tr\left[S_u(k_1+p_2)\gamma_5 S_d(k_1+p_1)\gamma_5 S_s(k_1+k_2)
\gamma_5\right]
\equiv 0.
\ena
In  Ref.~\cite{Cheng:1992ff} it was shown that
the vanishing strong couping for $\Xi_c^0\to\Xi^{+}_c\,\pi^-$
transition is a consequence of heavy quark and chiral symmetries.
Hence it is a model-independent statement.  
Here, one has to comment that there are two kinds of the interpolating
currents for the $\Lambda$-type baryons $(\Lambda_Q,\Xi_Q)$ where $Q=b,c$.
They are written as 
$\epsilon^{abc} Q^a(u^b C\gamma_5 s^c)$ (scalar diquark) and
$\epsilon^{abc}\gamma_\alpha Q^a(u^b C\gamma^\alpha\gamma_5 s^c)$ (vector diquark).
For the details, see
Refs~\cite{Shuryak:1981fza,Grozin:1992td,Groote:1996xb,Ivanov:1996fj}.

It is widely accepted that S-wave amplitude is saturated by the $1/2^-$
resonances, see, e.g., Refs~\cite{Marshak,Bailin:1977gv} for the
original suggestions and
\cite{Ebert:1983yh,Ebert:1983ih,Cheng:1985dw,Cheng:2018hwl}
for the subsiquent applications. Ordinarily, their contributions are calculated
by using the well-known soft-pion theorem in the current-algebra approach.
It allows one to express the parity-violating S-wave amplitude in terms of
parity-conserving matrix elements. In our case, one has
\bea
A_{1/2^-}\left(\Xi_c^0\to \Lambda_c^+ + \pi^-\right)
&=&
\frac{1}{f_\pi} A_{\Xi^+_c\Lambda_c^+}, 
\nn
A_{\Xi^+_c\Lambda_c^+}&=&
  V_{\rm CKM}^{(u)} \left(C_2^{(u)}-C_1^{(u)}\right) a^{(u)}_{\,\Xi^{+}_c\,\Lambda^+_c}
+ V_{\rm CKM}^{(c)} \left(C_2^{(c)}-C_1^{(c)}\right) a^{(c)}_{\,\Xi^{+}_c\,\Lambda^+_c}.
\label{eq:1/2minus}
\ena
The quantities $a^{(u,c)}_{\,\Xi^{+}_c\,\Lambda^+_c}$ and
$b^{(u,c)}_{\,\Xi^{+}_c\,\Lambda^+_c}$  are defined by
Eqs.~(\ref{eq:Xicp-weak-u})-(\ref{eq:Xicp-weak}).

Finally, the transition $\Xi_c^0\to \Lambda_c^+ + \pi^-$ amplitude
is written in terms of invariant amplitudes as
\be
<\Lambda_c^+\,\pi^-| {\mathcal H}_{\rm eff} | \Xi_c^0 >
= \frac{G_F}{\sqrt{2}}\, \bar u(p_2) \left(A+\gamma_5 B\right) u(p_1)
\label{eq:ampl}
\en
where $A$ and $B$ are given by
\bea
A &=& A_{\rm SD} + A_{\rm LD}, \qquad
A_{\rm LD} = A_{\Sigma^0_c} + A_{\Xi^{'\,+}_c}  + A_{1/2^-} ,
\nn
B &=& B_{\rm SD} + B_{\rm LD}, \qquad
B_{\rm LD} = B_{\Sigma^0_c} + B_{\Xi^{'\,+}_c}.
\label{eq:fullAB}
\ena

It is more convenient to use helicity amplitudes $ H_{\lambda_1 \lambda_M}$
instead of invariant ones $A$ and $B$ as described in~\cite{Korner:1992wi}.
One has 
\be
H^V_{\tfrac12\,t} = \sqrt{Q_+}\,A\,,\qquad H^A_{\tfrac12\,t} = \sqrt{Q_-}\,B\,,
\en 
where $m_\pm=m_1\pm m_2$, $Q_{\pm}=m_\pm^2-q^2$.

Finally, the two-body decay width reads
\be
\Gamma(B_1\to B_2+M) = 
\frac{G_F^2}{32\pi}\frac{\mathbf{|p_2|}}{m_1^2}\,{\mathcal H}_S\,,
\quad
{\mathcal H}_S =  2\Big(\Big|H^V_{ \tfrac12\,t}\Big|^2 \,+\, 
\Big|H^A_{\tfrac12\,t}\Big|^2\Big) \, 
\label{eq:width}
\en
where $\mathbf{|p_2|}=\lambda^{1/2}(m_1^2,m_2^2,q^2)/(2m_1)$.

\section{Numerical results}
\label{sec:results}

Our covariant constituent quark model
contains a number of model parameters which have been determined by
a global fit to  a multitude of decay processes.  The values of the constituent
quark masses $m_q$ are taken from the last fit in~\cite{Gutsche:2015mxa}.
In the fit, the infrared cutoff parameter $\lambda$ of the model
has been kept fixed as found in the original paper~\cite{Branz:2009cd}.
Table~\ref{tab: fitmas} shows as below: The size parameters of light meson
were fixed by fitting the data on the leptonic decay constant.
The numerical values of the size parameters and the leptonic decay constants
for pion is shown in Table~\ref{tab:pion}.
\begin{table}[H]
	\centering
	\def\arraystretch{1.1}
\caption{Constituent quark masses and infrared cutoff parameter $\lambda$.}
	\label{tab: fitmas}       
\vskip 2mm        
	\begin{tabular}{ccccc}
\hline\hline          
$m_{u/d}$        &      $m_s$     &      $m_c$       &   $\lambda$    &
\\\hline
\hspace*{1em} 0.241 \hspace*{1em} & \hspace*{1em} 0.428 \hspace*{1em} &
\hspace*{1em} 2.16\hspace*{1em}  & \hspace*{1em} 0.181 \hspace*{1em}  &
\hspace*{1em} {\rm GeV} \\
\hline\hline
	\end{tabular}
\end{table}
\begin{table}[H]
	\centering
	\def\arraystretch{1.1}
\caption{Size parameter and leptonic decay constant of pion.}
	\label{tab:pion}       
\vskip 2mm        
	\begin{tabular}{cccc}
\hline\hline 
\def\arraystretch{1.3}
  \hspace*{1em}  \text{Meson} \hspace*{1em} &  \hspace*{1em}
  $\Lambda_M$ \text{(GeV)}\hspace*{1em}  &\hspace*{1em} $f_M$ \text{(MeV)}
  \hspace*{1em}
  & \hspace*{1em}  $f_M^{\rm expt}$ \text{(MeV)}\hspace*{1em} \\
\hline
\qquad 
\text{Pion}    & 0.871    &   130.3    & 130.41 $\pm$ 0.20   \\
\hline\hline 
\end{tabular}
\end{table}
Since the experimental data of the single charm baryon decays become to appear
recently, we will assume for the time being that the size parameters of
all single charm baryons are the same. 
In Fig.~\ref{fig:Lambda} we plot the dependence on this parameter
denoted as $\Lambda_c$ of branching fractions $\Xi_c^0 \to \Lambda_c^+ + \pi^-$.
One can see that  the measured branching fraction can be
accommodated in the framework of this work by having
$\Lambda_c\approx 0.61$~GeV. In addition to the line describing
the central value of the experimental data, we also display the strip
corresponding to experimental uncertainties. In order to estimate the
uncertainty caused by the choice of the size parameter we allow the size
parameter to vary from $\Lambda_{c\,\rm min}=0.54$ to
$\Lambda_{c\,\rm max}=0.66$~GeV that correspond to the intersections of
the theoretical curve for branching fraction with the experimantal
lower and upper error bars.

We evaluate the mean $\bar\Gamma=\sum_{i=1}^N\Gamma_i/N$ and the mean
square deviation \newline
$\sigma^2=\sum_{i=1}^N(\Gamma_i-\bar\Gamma)^2/N$.
Finally, our result for the branching fraction reads as
\be
   {\mathcal B}\left(\Xi_c^0 \to \Lambda_c^+ + \pi^- \right)
   = (0.54 \pm 0.11)\%
   \en
   which should be compared with the data from LHCb and Belle:
   ${\mathcal B}=(0.55\pm0.02\pm0.18)\%$~\cite{LHCb:2020gge} and 
   ${\mathcal B}=(0.54\pm0.05\pm0.12)\%$~\cite{Belle:2022kqi}.

For comparison, we plot in Fig.~\ref{fig:Lambda} both the separate
SD-contributions coming from the diagrams with topologies
Ia, IIa, IIb, and III and the LD-contributions coming from the
pole diagrams. It is readily seen that the SD-contributions are much smaller
than those coming from the pole LD-diagrams.
\begin{figure}[H]
	\centering
	\def\arraystretch{1}
	\begin{tabular}{c}
		\includegraphics[width=\textwidth]{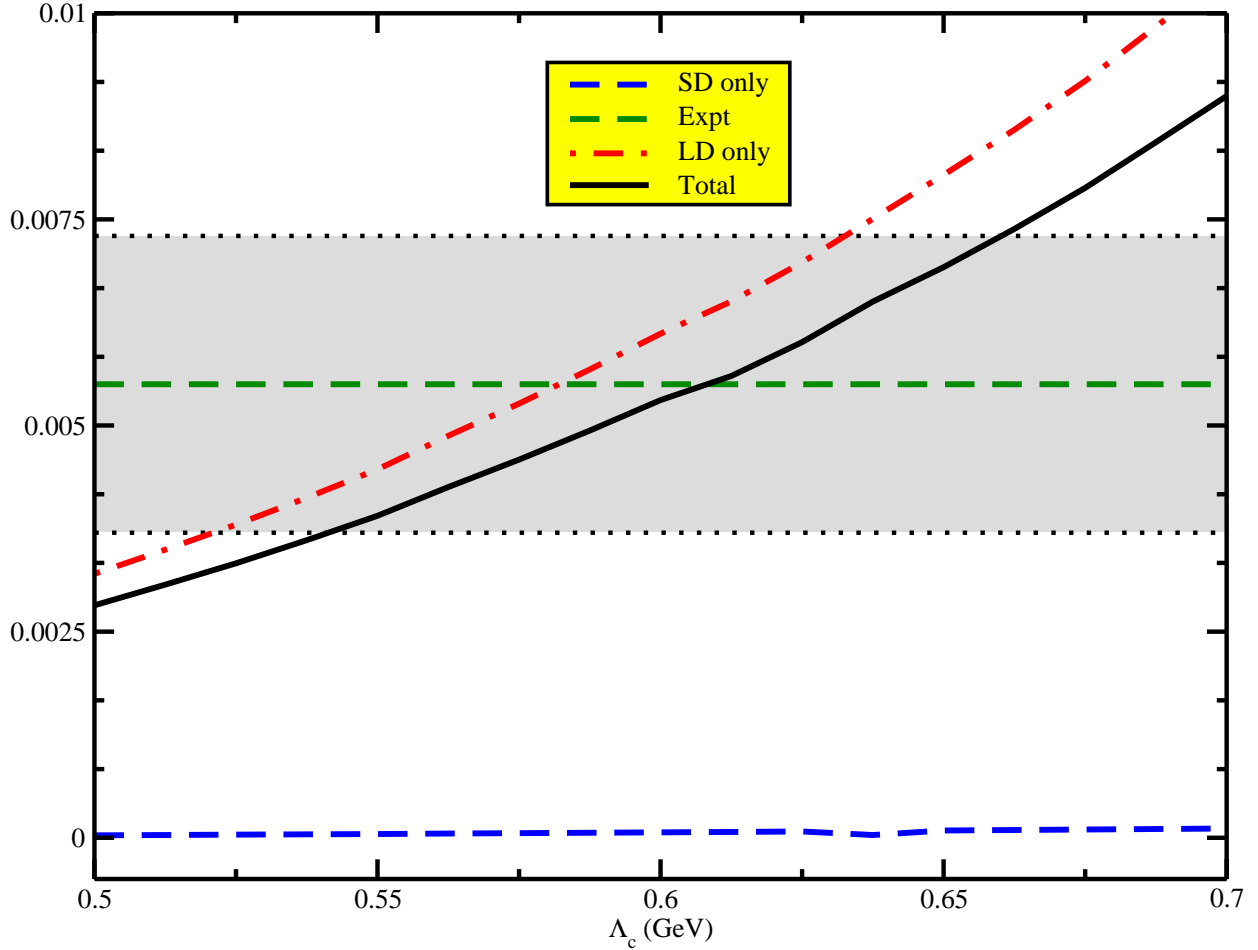}
	\end{tabular}
	\caption{Dependence of the branching fractions
		on the size parameter.}
	\label{fig:Lambda}
\end{figure}
The numerical results for the SD, LD and full amplitudes are shown
in Table~\ref{tab:ampl}. One can see that $|A_{\rm LD}|> |A_{\rm SD}|$.
\begin{table}[H]
  \caption{SD, LD and full amplitudes in units of GeV$^2$.}\vskip 3mm
	\centering
	\def\arraystretch{1.2} 
	\begin{tabular}{|c|c|c|c|}
\hline
\qquad Amplitudes \hspace*{2em} &\qquad SD  \hspace*{2em} &
\qquad LD \hspace*{2em}  & \qquad SD+LD  \hspace*{2em}  \\
\hline
A-ampl.    & 0.0156  & -0.0751 & -0.0595
\\                
B-ampl.    & 0.166   & -5.378  & -5.212
\\
\hline
\end{tabular}
\label{tab:ampl}
\end{table}	
   
Also it would be instructive to evaluate the asymmetry parameter
defined by 
\be
\alpha = \frac{|H_{1/2 \, t}|^2-|H_{-1/2 \, t}|^2}{|H_{1/2 \, t}|^2+|H_{-1/2 \, t}|^2}
= - \frac{2 \kappa A B}{A^2 +\kappa^2 B^2},
\label{eq:asym}
\en
where $\kappa = \bm{|{\rm p}_2|}/(E_2 + m_2)$ and
$E_2 = (m_1^2+m_2^2-q^2)/(2 m_1)$.
The numerical value of the asymmetry parameter is found to be equal to
\be
\alpha = -0.751.
\label{eq:asym-value}
\en

Finally, we compare our results obtained for the branching fraction
and the asymmetry parameter with other the data and other approaches
in Table~\ref{tab:comparison}.

\begin{table}[H]
	\caption{Comparison of our findings with other approaches.}\vskip 3mm
\centering
\def\arraystretch{1.3} 
\begin{tabular}{|c|c|c|}
\hline
Approach \quad 	& \hspace*{1cm} BR($\Xi_c^0 \to \Lambda^+_c \pi^-$)\%
\hspace*{1cm} & \,\, Asymmetry \,\,  \\
\hline
Our model
 &\,\, $0.54\pm 0.11$ &  $-0.75$ \\
\hline
LHCb~\cite{LHCb:2020gge}   &\,\, $ 0.55\pm 0.02\pm 0.1  $ & $\ldots$ \\
\hline
Belle~\cite{Belle:2022kqi} &\,\, $ 0.54 \pm 0.05 \pm 0.12 $ & $\ldots$ \\
\hline
Voloshin~\cite{Voloshin:2019ngb} &\,\, $ > 0.025\pm 0.015 $ & $\ldots$ \\
\hline
Gronau and Rosner~\cite{Gronau:2016xiq} (construc) &\,\,$  0.194 \pm 0.070 $
&  $\ldots$  \\
\hline
Gronau and Rosner~\cite{Gronau:2016xiq} (destruc) &\,\, $  < 0.01 $ & $\ldots$
\\
\hline
Faller and Mannel~\cite{Faller:2015oma} &\,\, $  < 0.39 $  & $\ldots$ \\
\hline
Cheng et al.~\cite{Cheng:2022jbr} &\,\, $  0.72 \pm 0.07 $ & $0.46 \pm 0.05 $\\
\hline
Niu et al.~\cite{Niu:2021qcc} &\,\, $  0.58 \pm 0.21 $ &  $ -0.16 $ \\
\hline
\end{tabular}
\label{tab:comparison}
\end{table}

\section{Summary and conclusion}
\label{sec:summary}

We have studied two-body nonleptonic $\Delta C=0$  decay  
$\Xi^0_c\to \Lambda_c^++\pi^-$ in the framework of the covariant confined quark
model (CCQM) with account for both  short and long distance effects. 
The short distance effects are induced by four topologies of external and
internal weak $W$-interactions, while long distance effects are saturated
by an inclusion of the so-called pole diagrams. Pole diagrams are generated
by resonance contributions of the low-lying spin $\frac12^+$
($\Sigma_c^0$ and $\Xi_c^{\,\prime\,+}$) and spin $\frac12^-$ baryons. The last
contributions are calculated by using the well-known soft-pion theorem.
It is found that the contribution of the SD diagrams is significantly
suppressed, by more than one order of magnitude in comparison with data.
The most significant contributions are coming from the intermediate
$\frac12^+$ and $\frac12^-$ resonances. We can get consistency with
the experimental data for the value of size parameter being equal to
$\Lambda\approx 0.61$~GeV.

\begin{acknowledgments}

The research  has been funded by the Science Committee of
the Ministry of Science and Higher Education of the Republic
of Kazakhstan (Grant No. AP19678771).
V.E.L. acknowledges the support 
by  ANID PIA/APOYO AFB220004 (Chile),
by FONDECYT (Chile) under Grant No. 1230160,
and by ANID$-$Millen\-nium Program$-$ICN2019\_044 (Chile).

\end{acknowledgments}

\end{document}